# Study of levitating nanoparticles using ultracold neutrons

V.V. Nesvizhevsky [1], A.Yu. Voronin [2], A. Lambrecht [3], S. Reynaud [3]

[1] Institut Laue-Langevin, Grenoble, France, nesvizhevsky@ill.eu
[2] Lebedev Institute, Moscow, Russia
[3] Laboratoire Kastler Brossel, Paris, France

**Abstract**

Physical adsorption of atoms, molecules and clusters on surface is well known. It is linked to many other phenomena in physics, chemistry, biology, and has numerous practical applications. Due to limitations of analytical tools usually used, studies of adsorption are limited to the particle sizes of up to $\sim 10^2$-$10^3$ atoms. Following a general formalism developed in this field we apply it to even larger objects and discover qualitatively new phenomena. A large particle is bound to surface in a deep and broad potential well formed by van der Waals/ Casimir-Polder forces (vdW/CP) appearing due to the particle and surface electric polarization. The well depth is significantly larger than the characteristic energy $\frac{3}{2}k_B T$ of a particle thermal motion; thus such a nanoparticle is settled in long-living states. Nanoparticles in high-excited states form two-dimensional gas of objects bound to surface but quasi-freely traveling along surface at certain conditions. A particularly interesting feature of this model consists in prediction of small-energy-transfer inelastic scattering of ultracold neutrons (UCN) on solid/ liquid surfaces covered by such levitating nanoparticles/ nano-droplets. The change in UCN energy is due to the Doppler shift induced by UCN collisions with nanoparticles/ nano-droplets; the energy change is about as small as the UCN initial energy. We compare theoretical estimations of our model to all existing data on inelastic scattering of UCN with small energy transfers and state that they agree quite well. As our theoretical formalism provides robust predictions of some data and the experimental data are rather detailed and precise, we conclude that the recently discovered intriguing phenomenon of small heating of UCN in traps is due to their collisions with such levitating nanoparticles. Moreover, this new phenomenon might be relevant to the striking contradiction between results of the neutron lifetime measurements with smallest reported uncertainties as it might cause major false effects in these experiments; thus it affects fundamental conclusions concerning precision checks of unitarity of the Cabibbo-Kobayashi-Maskawa matrix, cosmology, astrophysics. Dedicated measurements of UCN inelastic scattering on specially prepared surfaces and nanoparticles/ nano-droplets levitating above them, might provide a unique method to study surface potentials. The present study had been performed prior to the experiment presented in ref. (1), and strongly motivated this experiment. The new data and their analysis will be published in detail elsewhere.

1. Introduction

An atom above a surface is affected by a force originating from polarization of this atom and atoms in the surface. This interaction is attractive on a nanometer scale (van der Waals interaction is dominant at smaller distances, Casimir-Polder interaction is more important on larger distances, vdW/CP); it is repulsive on Angstrom scale (reflection due to overlapping of the electron clouds from the probe atom and the surface atoms). The resulting potential well has a minimum at a distance of about a nanometer to surface; the potential flattens at larger distances to give a constant value at infinity. The distance of "zero force" (the minimum in the potential well) is defined by balance between the two forces: the weaker is attraction, the larger is this characteristic distance.

To remind known facts: surface diffusion of adsorbed atoms, molecules and clusters is studied in an extremely broad range of sciences including surface physics and chemistry (2), (3). Various analytical tools may be used to study it, such as field ion microscopy and scanning tunneling microscopy. Due to experimental constraints most studies are limited to well below the substrate melting point (4). Surface diffusion rates and mechanisms are strongly affected by a variety of factors. Surface diffusion is an important concept in surface phase formation, epitaxial growth, heterogeneous catalysis, and other topics in surface science (5). Applications include those in the chemical production and semiconductor industries, catalytic converters, integrated circuits used in electronic devices, silver halide salt used in photographic films (5).

Due to limitations of analytical tools of studying the adsorption particles, these studies are limited to particles of relatively small size of up to $\sim 10^2$-$10^3$ atoms. Following the general formalism developed in the field, we apply the theoretical analysis to even larger particles and discover qualitatively new phenomena related to larger sizes (masses) of particles involved. In fact, the depth of the potential well affecting motion of a particle near a surface is proportional, if the particle is small enough, to the number of atoms in it, while the thermal energy attributed to the particle is always $\sim \frac{3}{2}k_B T$. As a result, a particle with a size of a few nanometers is deeply bound in the potential well with the depth of significantly larger than $\frac{3}{2}k_B T$. On the other hand, its motion along a sufficiently flat and uniform surface is not strongly affected by the bounding potential. Thus such nanoparticles might form two-dimensional gas of objects moving quasi-freely along a surface and bound in the direction perpendicular to it. Due to the fact that their size is much larger than a typical inter-atomic distance, also due to relatively large mean distance of such "levitating" nanoparticles to surface, of the order of a nanometer, their interaction with a surface is averaged over a huge number of atoms. Therefore it might be not that much affected by impurities in and roughness of a surface; thus a predictive power of our model is quite high in this case. Our theoretical formalism describes states of nanoparticles in the potential well formed by the vdW/CP forces on one side and by quantum reflection from surface on another side. We do not consider here more general cases of chemisorption and electrostatic interactions when a particle or/and a surface is electrically charged. As will be clear from the analysis below, in the cases of chemisorption and attractive electrostatic interaction, nanoparticles are too strongly bound to surface, thus they do not contribute significantly to the inelastic scattering of UCN. In the case of a repulsive electrostatic interaction, nanoparticles will be bound at even larger mean distances to surface, and to each other, or not bound at all, thus they would contribute in a way similar to that described in the present article, thus not requiring a particular treatment at the present level of precision. It is curious to note that the present model has been motivated in a significant extend by other phenomena of quantum levitation of massive particles above surface that we have been considering previously (6), (7), (8), (9).

We argue that such levitating nanoparticles have been already observed in the experiments using storage of ultracold neutrons (UCN) (10), (11), (12) in closed traps. We will describe the interaction of UCN with levitating nanoparticles and show that nanoparticles behave as "free" objects in a way similar to that considered in refs. (13), (14). In other words, we deal essentially with the Doppler shift in neutron energy due to quasi-classical billiard-ball collisions of an UCN with a moving nanoparticle; a more complex case of a very rough surface, which bounds nanoparticles in more-than-one direction, will be considered elsewhere. The phenomenon of small-energy-transfer inelastic scattering of UCN on surfaces was discovered in a series of experiments studying UCN storage in material traps (see (15), (16), (17), (18), (19) and further publications). We call this phenomenon the small heating of UCN (an increase in energy is significantly more probable than a decrease in energy); and we call such up-scattered neutrons VUCN (Vaporizing UCN – in analogy to vaporization of molecules from liquid or solid surfaces).

The first measurements were followed by other studies (20), (21), (22), (23), (24), (25), (26), (27), (28), (29), (30) carried out by several research groups, which essentially confirmed the initial observation but suggested controversial estimations of the absolute VUCN production rates. This controversy is explained by several factors: a) measurements of such a kind, involving low-probable effects, require careful control of many systematic effects, which could imitate or hide the VUCN production; b) the rate of VUCN detection depends strongly on the ranges (windows) of initial and final energies of UCN, which have not been under control in some experiments, c) as we will see from analyzes of the spectra presented in this work, very high energy resolution is needed; it has been provided so far only in the BGS spectrometer (27), (30). Thus the probability of VUCN production per UCN-wall collision might be largely miss-estimated in other experiments; d) the VUCN production probability depends strongly on procedure of a sample (a surface) treatment or preparation, as this has been demonstrated, for instance, in ref. (27). Therefore a direct comparison of the absolute probability of VUCN production on different samples is meaningless. In particular, it depends strongly on the temperature of solid sample heating during and before measurements, or on vapor pressure above liquid surfaces and the amount and composition of gases diluted in a liquid.

To summarize the status of all these studies, trap walls (samples) of two kinds were used in these experiments: liquid and solid walls (samples). While the origin of small heating of UCN on liquid surfaces has been still under discussion (31), (22), (20), (21), (32), (33), (34), (13) (see also (35), (36), which rule out some hypotheses), the origin of small heating of UCN on solid surfaces is usually attributed to weakly bound nanoparticles on surface (13), (30), (37); note that increase in the scattering cross-section due to clustering of atoms was noted already in ref. (34). Here we will compare the theoretical estimations within our model to all precision experimental data available on inelastic scattering of UCN on nanoparticles bound to solid surfaces (30), (37). The experimental data include the following measurements: 1) The probability of inelastic scattering and the spectrum of UCN scattered on detonation diamond nanoparticles above a copper surface as a function of temperature (liquid nitrogen and ambient temperature); 2) The probability of inelastic scattering and the spectra of UCN scattered on nanoparticles formed due to thermal diffusion on metallic surfaces (copper, stainless steel, beryllium) as a function of temperature (liquid nitrogen and ambient temperature), also as a function of the temperature of surface preceding treatment. 3) Absence of small-energy-transfer inelastic scattering of UCN on sapphire monocrystal surface, which could be considered as a "zero test" of validity of our model.

A scheme of the gravitational spectrometer used in measurements of inelastic scattering of UCN (27), (30) is shown in Fig. 1; principles of its operation are explained in detail in the references therein. The spectrometer allows us to measure the probability of inelastic scattering of UCN from a broad initial energy window to another broad window of final UCN energies. The efficiency of detection of produced VUCN was measured experimentally as a function of VUCN energy; it changes nearly linearly from ~0.5 at small energy transfers of ~10neV to zero at large energy transfers of ~160neV. A dead energy zone of ~10neV separates the highest energy in the initial UCN spectrum from the smallest energy in the spectrum of detected VUCN. Various samples could be installed to the bottom of the gravitational spectrometer as shown in Fig. 1.

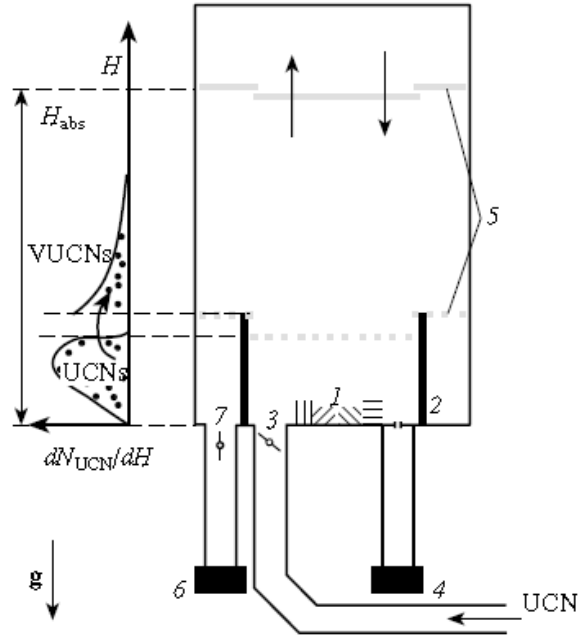

*Fig. 1. A layout of the gravitational spectrometer of total UCN energy: (1) sample, (2) gravitational barrier, (3) entrance valve, (4) monitor detector, (5) UCN absorber, (6) UCN detector, and (7) exit valve. A principle scheme of a procedure to measure small energy transfers is illustrated in the insert on left. From bottom to top: an initial UCN spectrum, a dead-zone insensitive to VUCN, and an energy window of efficient VUCN detection.*

In Section 2 we analyze shapes of the bounding potential wells formed by quantum reflection on one side and by vdW/CP attraction on another side. Also we describe quantum states of a nanoparticle in such a potential well. The characteristic ranges of the problem parameters are estimated. In Section 3 we develop a formalism describing scattering of UCN on levitating nanoparticles. In particular, we show that nanoparticles could be considered, at some conditions, as freely moving objects. Further we consider conditions under which UCN scattering on levitating nanoparticles would be affected by the bounding potential shape. The characteristic parameters of the problem are presented for both limiting cases. Section 4 is devoted to describing the model of two-dimensional gas of freely levitating nanoparticles. In particular, the characteristic ranges of parameters of the problem are estimated. In Section 5 we compare our theoretical estimations to all experimental data on inelastic scattering of UCN available. The data include measurements with nanoparticles levitating above solid surfaces (30), (37), in particular so-called "temperature resonance": sharp increase in the probability of inelastic scattering of UCN on stainless steel surface as a function of the temperature of sample preceding heating (30). We argue that inelastic scattering of UCN on solid and liquid surfaces is due to the Doppler shift in their energy following from their collisions with levitating nanoparticles. In Section 6 we list various manifestations of the phenomenon of quantum levitation of nanoparticles in UCN storage experiments including: its relevance to the recent discovery of small-energy-transfer inelastic scattering of UCN on solid and liquid surfaces, its eventual relevance to so-called anomalous losses of UCN (38), its relevance to unnoticed false effects in the precision neutron lifetime experiments (39), (40), (41), (42) affecting fundamental conclusions, in particular on unitarity of the Cabibbo-Kobayashi-Maskawa matrix (43), (44) in cosmology and astrophysics (45). In dedicated experiments on inelastic scattering of UCN, we would like to study shapes of the potential wells bounding levitating nanoparticles.

## 2. Bounding potential well

*2.1. Shape of the bounding potential well.*

In this subsection, we discuss the shape of a potential well associated with the vdW/CP forces between a flat surface and a nanoparticle. The names in the vdW/CP abbreviation stand for the same dispersive interaction between two objects coupled to vacuum fluctuations. The potential might be evaluated using a general expression involving only the scattering properties of the two objects (46). This expression is discussed, for example, in (47) for an atom above a plane, and in (48), (49), (50) for a metallic nanosphere above a plane. We present here main features of this expression, which are of importance for the forthcoming discussion; more detailed derivations are presented elsewhere (51). We consider here, in particular, an example of a diamond nanosphere above a copper plane, which is of relevance to measurements of inelastic scattering of UCN in (30).

In order to calculate the vdW/CP energy in the scattering approach, first, we have to describe the scattering properties of the two objects. For the copper plane, scattering is characterized by Fresnel reflection amplitudes, the form of which is fixed by the dielectric response function $\varepsilon$, or the index of refraction $n$. Here we use a simple Drude model for describing the metallic response of copper bulk:

$$\varepsilon(\omega) = n^2 = 1 - \frac{\omega_P^2}{\omega(\omega + i\gamma)}, \tag{2.1}$$

where $\omega_P$ is the plasma frequency proportional to the density of conduction electrons in the metal, $\gamma$ is the damping rate, which measures the relaxation of these electrons, and $\omega$ is the frequency of field fluctuations under study. For frequencies much lower than $\omega_P$, or for wavelengths much larger than the plasma wavelength $\lambda_P = 2\pi c/\omega_P$, $\varepsilon(\omega)$ is very large, which means that copper tends to behave as a good reflector. For larger frequencies or smaller wavelengths, in contrast, the reflection properties are poorer. The transition between these two regimes will manifest in the vdW/CP energy when the distance between two objects approaches the plasma wavelength ($\lambda_P^{Cu} = 136nm$ for copper). Relaxation parameter $\gamma$ is small when compared to $\omega_P$ for good metals ($\gamma \approx 0.0033\omega_P$ in copper). This implies that its influence would only be perceived at distances larger than those considered in the present study.

For the diamond nanosphere, scattering is characterized by Mie scattering amplitudes, the form of which is fixed by the dielectric response function $\varepsilon$, or the index of refraction $n$. Here we use a simple Sellmeier model for describing the dielectric response of diamond bulk:

$$\varepsilon(\omega) = n^2 = 1 + \sum_k \frac{b_k \omega_k^2}{\omega_k^2 - \omega^2}, \tag{2.2}$$

where $\omega_k$ are the resonance frequencies characterizing the material, and $\omega$ is the frequency of field fluctuations under study. Note that the damping rates are neglected here since they do not play any significant role for vdW/CP calculations. For frequencies much lower than $\omega_1$, that is also for wavelengths much larger than the resonance wavelength $\lambda_1 = 2\pi c/\omega_1$, $\varepsilon(\omega)$ tends to its static limit $\varepsilon(0) = n(0)^2 = 1 + \sum_k b_k$. For much larger frequencies or much smaller wavelengths, in contrast, the dielectric properties are poor. For diamond, a good description is obtained with a single component. The transition between these two regimes will affect the vdW/CP energy when the distance of approaching the two objects will be of the order of the resonance wavelength ($\lambda_1 \approx 106nm$ for diamond).

We then use the scattering approach as was done is (48), (49), (50) for the case of a nanosphere above a plane. We thus obtain the interaction energy $E$ as an integral of dephasings corresponding to all modes of electromagnetic vacuum. We do not repeat here these calculations, which can be found in (48), (49), (50) but only show the results in Fig. 2. Precisely, we plot the absolute value $|E|$ of the interaction energy ($E$ is negative) as a function of the distance $L$ of the closest approach of the sphere to the plane, for different values of the radius $R$. Logarithmic scales are used on abscissa as well as an ordinate axis. The figure shows clearly the transition between short and long distance behavior.

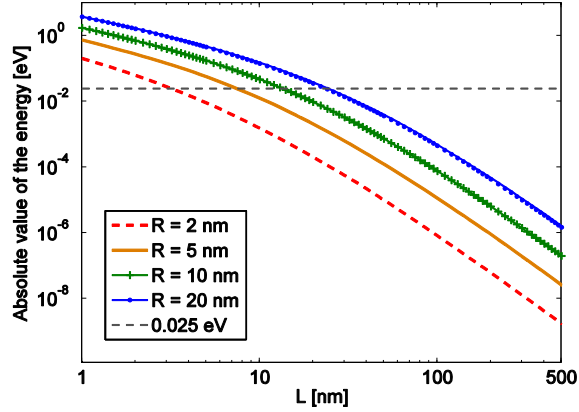

*Fig. 2. The absolute value of the vdW/CP interaction energy as a function of the distance of closest approach of the diamond sphere to the copper plane, for the sphere radii 2nm, 5nm, 10nm, and 20nm. The mean thermal energy $\frac{3}{2}k_BT$ (25meV at ambient temperature) is indicated as the horizontal dashed line.*

The long-distance behavior, above 100nm, corresponds to the Casimir-Polder (CP) limit. It might be recovered with the copper plane considered as perfectly reflecting and diamond as having a constant dielectric constant $\varepsilon(0)$. An approximation of the interaction energy is thus obtained as

$$V_{CP}(L) = -\frac{4\pi C_4 R^3}{3L^2(L+2R)^2}, \quad C_4 = \frac{9\hbar c \alpha_0}{32\pi^2}, \quad \alpha_0 = \frac{n_0^2-1}{n_0^2+2}. \tag{2.3}$$

The short distance behavior, below 30nm, corresponds to van der Waals (vdW) limit. It might be recovered as an integral over frequencies from zero to the cutoffs $\omega_P, \omega_1$. An approximation of the interaction energy is thus obtained as

$$V_{vdW}(L) = -\pi C_3 \left(\frac{2R(L+R)}{L(L+2R)} - Log\left(\frac{L+2R}{L}\right)\right), \quad C_3 = \frac{3\hbar c \alpha_0}{16(\sqrt{2}\lambda_P + \sqrt{1-\alpha_0}\lambda_1)}. \tag{2.4}$$

In the following, we will use the full expression of the interaction energy (drawn in Fig. 2) rather than the approximations (2.3) and (2.4). We also show the mean thermal energy $\frac{3}{2}k_BT \approx 25meV$ (at ambient temperature) as the horizontal dashed line in Fig. 2.

The repulsive potential at close-to-contact distances of a few Angstroms might be modeled using the Lennard-Jones potential, which increases very rapidly with decreasing the distance between the nanoparticle and the surface. When considering realistic shapes of solid surfaces (not flat) and nanoparticles (not spherical), one might take into account that a close contact is achieved in a few points only. Thus a few atoms in the nanoparticle participate in the repulsion, while all atoms in it contribute to the attraction. These effects are negligible for liquid surfaces and nano-droplets. The surface/nanoparticle roughness effects will be taken into account phenomenologically in Section 2.2.

*2.2. Quantum states of nanoparticles in the vdW/CP potential well.*

Consider a nanoparticle above a rough surface, with the roughness amplitude much smaller than the roughness correlation length; this approximation is valid for all surfaces used in the neutron experiments considered in the present study. We will show that a nanoparticle will be bound, at certain conditions, in a one-dimensional potential well (in the direction perpendicular to the surface) formed by the vdW/CP interactions; it will move quasi-freely in two other dimensions along the surface. We assume that nanoparticles in the one-dimensional well are distributed according to the general statistical law, i.e. the number of nanoparticles $N_i$ at a given energy level $E_i$ equals

$$N_i = \frac{N}{Z} exp\left(-\frac{E_i}{k_B T}\right), \tag{2.5}$$

$$Z = \sum_i exp\left(-\frac{E_i}{k_B T}\right) \approx \int_{E_0}^{\infty} exp\left(-\frac{E}{k_B T}\right)\frac{dn}{dE}(E)dE. \tag{2.6}$$

Here $\frac{dn}{dE}(E)$ is the level density for a given energy $E$, $E_0 < 0$ is the ground state energy in the vdW/CP potential well. The value of $E_0$ is involved in calculation of the statistical sum $Z$ thus in calculation of physical values. In a turn, one has to know the nanoparticle-surface interaction at smallest distances in order to calculate $E_0$. As a nanoparticle cannot approach surface closer than a characteristic roughness/surface diffuseness scale, we will use the following regularizing procedure:

$$V_{vdW}(L) = \begin{cases} -\pi C_3 \left( \frac{2R(L+R)}{L(L+2R)} - Log\left(\frac{L+2R}{L}\right) \right), L_c < L < \lambda_P \\ -\pi C_3 \left( \frac{2R(L_c+R)}{L_c(L_c+2R)} - Log\left(\frac{L_c+2R}{L_c}\right) \right), L < L_C \end{cases}. \quad (2.7)$$

Here $L_c$ is a free parameter of the order of a typical surface roughness/diffuseness of the surface potential.

Profiting from the fact that the nanoparticle mass is large, we get for the ground energy $E_0 \approx V_{vdW}(L_c)$. For instance, if $L_c = 1nm$ and a diamond nanoparticle radius is $R = 4nm$ then the ground energy equals $E_0 \approx -1eV$; if $R = 8nm$ then $E_0 \approx -2.5eV$. Note that these values are significantly larger than the thermal motion energy, and would be even higher for an ideally flat surface.

A large nanoparticle mass, thus a large number of quantum states in the vdW/CP potential well, allows us to use the semi-classical Born-Somerfield quantization rule for calculating the quantum state energies:

$$\int_0^{L_t} \sqrt{2M(E_n - V_{vdW}(x))} dx = \pi n, \quad (2.8)$$

$$\frac{1}{\pi} \int_0^{L_t} \sqrt{\frac{M}{2(E_n - V_{vW}(x))}} dx = \frac{dn}{dE}, \quad (2.9)$$

$$E - V_{vdW}(L_t) = 0. \quad (2.10)$$

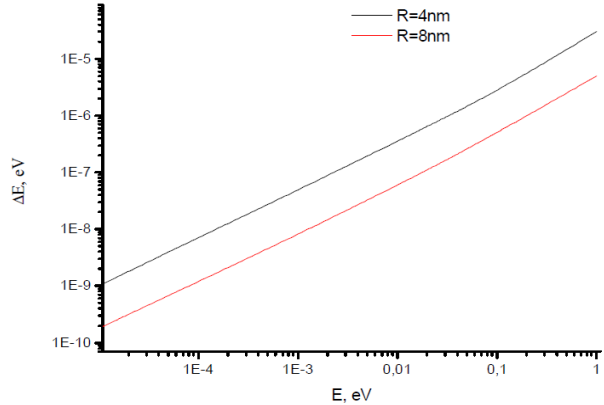

*Fig. 3 presents the level energy spacing as a function of the binding energy modulus for a diamond nanoparticle with the radii $R = 4nm$ and $R = 8nm$. An important feature of this spectrum consists in the fact that the level spacing is much smaller than the binding energy even for the lowest quantum states; also it is much smaller than $\sim \frac{3}{2} k_B T$ at ambient temperature.*

In the above formulas $L_t$ is the classical turning point, determined in eq. (2.10). The semi classical character of the nanoparticle motion suggests that the classical statistics can be applied to the nanoparticle gas. In this case the spatial and momentum distributions are separated:

$$n(x,p) = n_0 exp\left(-\frac{V_{vdW}(x)}{k_B T}\right) exp\left(-\frac{p^2}{2M}\right). \quad (2.11)$$

This means that nanoparticles are spatially distributed in accordance with the Boltzmann formula as a function of the distance to surface, while their momentum distribution is given independently by Maxwellian factor.

## 3. Interaction of UCN with levitating nanoparticles

To start with, let's consider the simplest case of freely levitating nanoparticles and show that this simple model describes well the data available.

*3.1. The cross-section of inelastic scattering of UCN on nanoparticles*

For completeness we reproduce here general results (52), which could be applied to calculate the energy transfer cross-section $\frac{d\sigma(\vec{V},\vec{v_n})}{dE}$ as a function of the nanoparticle velocity $\vec{V}$ and the neutron velocity $\vec{v_n}$ in the laboratory reference system. In the following we will take into account that the nanoparticle mass $M$ is much larger than the neutron mass $m$, thus the nanoparticle velocity in the laboratory reference system coincides with the velocity of the neutron-nanoparticle center-of-mass.

The energy transferred in a neutron-nanoparticle collision equals in the laboratory reference system to:

$$E = \vec{V}m(\vec{v_n}' - \vec{v_n}). \tag{3.1}$$

Here $\vec{v_n}'$ is the neutron velocity after collision. Since $(\vec{v_n}' - \vec{v_n})$ is invariant under a reference frame transformation, it is convenient to perform calculations of the cross-sections in the center-of-mass reference system. The same quantity could be expressed in terms of the neutron momentum in the center-of-mass reference system:

$$E = Vk_0(\cos(\theta') - \cos(\theta)). \tag{3.2}$$

Here $k_0 = m|\vec{v_n} - \vec{V}|$ is the incident momentum absolute value in the center-of-mass reference system; the axis is chosen to be parallel to the nanoparticle momentum, so $\theta'$ and $\theta$ are the angles between the nanoparticle momentum and the neutron momentum vector in the center-of-mass reference system after and before collision.

The total (invariant) cross-section of neutron scattering on nanoparticle is:

$$\sigma = \int_{-\infty}^{\infty} \frac{d\sigma}{dE} dE = \int |f_{cm}(\theta,\theta',\varphi,\varphi')|^2 d\cos(\theta')d\varphi'. \tag{3.3}$$

Here $f_{cm}$ is the scattering amplitude in the center-of-mass reference system.

Differentiation of eq. (3.2) results to

$$d\cos(\theta') = \frac{dE}{k_0 V}. \tag{3.4}$$

Thus the total cross-section (eqs. (3.3), (3.4)) equals

$$\sigma = \int_{-\infty}^{\infty} \frac{d\sigma}{dE} dE = \int |f_{cm}(\theta,\theta',\varphi,\varphi')|^2 d\varphi' \frac{dE}{k_0 V}, \tag{3.5}$$

and the differential cross-section equals

$$\frac{d\sigma}{dE} = \frac{1}{k_0 V} \int |f_{cm}(\theta,\theta'(E),\varphi,\varphi')|^2 d\varphi'. \tag{3.6}$$

The above expression gives us the inelastic differential cross-section for fixed $\vec{V}$ and $\vec{v_n}$ values using the elastic differential scattering cross-section in the center-of-mass reference system. In eq. (3.6) $\theta'(E)$ could be calculated using eq. (3.2):

$$\cos(\theta') = \frac{E}{k_0 V} + \cos(\theta). \tag{3.7}$$

For the scattering angle $\chi$ between the neutron momenta in the center-of-mass reference system before and after collision one gets:

$$\cos(\chi) = \cos(\theta')\cos(\theta) + \sin(\theta')\sin(\theta)\cos(\varphi' - \varphi). \tag{3.8}$$

In particular, the Born amplitude is a function of the transferred momentum $q$ only:

$$q = 2k_0 sin\left(\frac{\chi}{2}\right) = k_0\sqrt{2(1 - cos(\theta)cos(\theta') - sin(\theta)sin(\theta')cos(\varphi' - \varphi))}. \qquad (3.9)$$

*3.2. Spectrum of inelastically scattered UCN*

We consider here UCN colliding with ideal gas of free nanoparticles with the Maxwellian velocity distribution; the initial velocity of neutrons $v_n$ is fixed; the angular distribution of neutrons is isotropic. The corresponding neutron momentum distribution is:

$$n(\vec{p}) = \frac{1}{4\pi k_n^2}\delta(k_n - |\vec{p}|), k_n = mv_n. \qquad (3.10)$$

We average the cross-section (3.6) over nanoparticle velocities and all possible directions:

$$\frac{d\bar{\sigma}}{dE} = \frac{1}{4\pi k_n^2}\left(\frac{M}{2\pi k_B T}\right)^2 \int exp\left(-\frac{MV^2}{2k_B T}\right)V^2 dV d\Omega_V \frac{|f_{cm}(\theta,E,\varphi,\varphi')|^2 d\varphi'}{Vk_0}\delta(k_n - p)p^2 dp d cos(\theta) d\varphi. \qquad (3.11)$$

It is convenient to integrate over the variable $\vec{k_0} = \vec{p} - m\vec{V}$ instead of $\vec{p}$:

$$\frac{d\bar{\sigma}}{dE} = \frac{1}{2k_n^2}\left(\frac{M}{2\pi k_B T}\right)^{\frac{3}{2}} \int exp\left(-\frac{MV^2}{2k_B T}\right)VdVd\Omega_V \xi^2(\theta, E)\,\delta\left(k_n - \sqrt{k_0^2 + m^2V^2 + k_0 mVx}\right) k_0 dk_0 dx,$$
with $x = cos(\theta)$ and $\xi^2(\theta, E) = \int_0^{2\pi}|f_{cm}(\theta, E, \varphi')|^2 d\varphi'.$ $\qquad (3.12)$

Integration over $x$ results in:

$$\frac{d\bar{\sigma}}{dE} = \frac{1}{2k_n^2}\left(\frac{M}{2\pi k_B T}\right)^{\frac{3}{2}} \int exp\left(-\frac{MV^2}{2k_B T}\right)VdVd\Omega_V \xi^2(\theta, E)\,\alpha(x_0)\frac{k_n}{mVk_0}k_0 dk_0. \qquad (3.13)$$

Here

$$\alpha(x_0) = \begin{cases} 1, |x_0| \leq 1 \\ 0, |x_0| > 1 \end{cases},$$
$$x_0 = \frac{k_n^2 - k_0^2 - m^2V^2}{2mVk_0}. \qquad (3.14)$$

Integrating over $k_0$ and $V$, we should take into account constraints following from eq. (3.7), namely:

$$\left|\frac{E}{k_0 V} + x_0\right| \leq 1. \qquad (3.15)$$

Finally we get the following expression:

$$\frac{d\bar{\sigma}}{dE} = \frac{1}{2mk_n}\left(\frac{M}{2\pi k_B T}\right)^{\frac{3}{2}} \int exp\left(-\frac{MV^2}{2k_B T}\right)dVd\Omega_V \xi^2(\theta, E)\,\alpha\big(x_0(k_0, V)\big)\beta\big(x_0(k_0, V)\big)dk_0. \qquad (3.16)$$

Here

$$\beta(x_0) = \begin{cases} 1, \left|\frac{E}{k_0 V} + x_0\right| \leq 1 \\ 0, \left|\frac{E}{k_0 V} + x_0\right| > 1 \end{cases}. \qquad (3.17)$$

The conditions (3.14), (3.17) restrict the kinematically allowed region in phase-space; they can be written as follows:

$$|mV - k_n| \leq k_0 \leq mV + k_n,$$
$$\left|\sqrt{k_n^2 + 2mE} - mV\right| \leq k_0 \leq \sqrt{k_n^2 + 2mE} + mV. \qquad (3.18)$$

To calculate the cross-section $\xi^2$ in the center-of-mass one could use the Born amplitude of UCN scattering on a spherical nanoparticle with the radius $R$ and the real part of the optical potential $U_0$:

$$f_{cm} = -\frac{2mR^3}{(qR)^2} U_0 \left( \frac{\hbar \sin\left(\frac{qR}{\hbar}\right)}{qR} - \cos\left(\frac{qR}{\hbar}\right) \right). \tag{3.19}$$

One can show that for $qR < 1$, the coherent interaction results in sharp R-dependence of the Born cross-section:

$$|f_{cm}|^2 \sim \frac{4U_0^2 m^2 R^6}{9\hbar^2}, \tag{3.20}$$

while for $qR > 1$ it drops down quickly.

### 4. Two-dimensional gas of freely levitating nanoparticles and VUCN.

*4.1. The characteristic ranges of parameters of the problem.*

In this section, we estimate the characteristic parameters of the problem, namely 1) the nanoparticle radii and 2) the spatial extension of quantum states. We consider here 1) diamond nanoparticles above copper surfaces (as the most precise data have been measured for this case (30)), 2) neutron spectra and VUCN detection efficiency are equivalent to those in ref. (30) (the neutron detection efficiency was measured in this experiment).

In order to estimate the nanoparticle radii range, we take into account the following conditions (following ref. (13)): 1) The nanoparticle radius should not be too small, otherwise the coherent interaction cross-section is too low (see eq. (3.20): $\frac{d\bar{\sigma}}{dE} \sim R^6$), and 2) The nanoparticle radius should not be too large, otherwise nanoparticles are too heavy; thus their thermal velocity is too small, the energy transfer is too small, and VUCN could not be detected in the window of efficient VUCN detection see eq. (3.18)). Relative probability of VUCN detection in the BGS spectrometer (30) is shown in Fig. 4 as a function of the nanoparticle radius. As one can see, only nanoparticles with a radius within a well-defined range of 7-12 nm could be efficiently observed by means of measuring VUCN in the BGS spectrometer. The maximum sensitivity is sharply defined at $R \sim 10$ nm. The range of nanoparticle radii, corresponding to the efficiency of VUCN detection higher than 10% (1%) of the maximum efficiency, is shown in Fig. 5 with vertical solid thick (thin) lines.

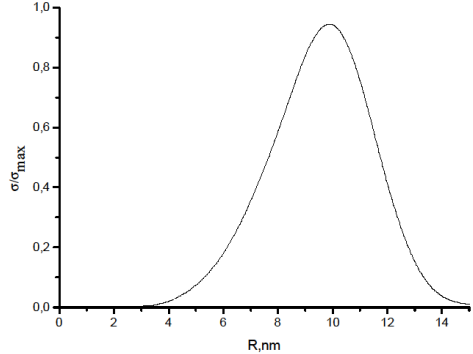

*Fig. 4. Relative probability of VUCN detection in the BGS spectrometer (the measured spectrometer efficiency $Eff(E)$ is taken into account) as a function of the diamond nanoparticle radius $R[nm]$.*

In order to estimate the spatial extension of quantum states along the direction perpendicular to the surface, we will define the cut-off parameter as follows: Nanoparticles should be bound by the vdW/CP potential well below some characteristic energy expressed in units of the thermal motion energy $\frac{3}{2} k_B T$. The classical turning point for a nanoparticle with the energy $-3k_B T$ ($-k_B T$) in the bounding vdW/CP potential is shown in Fig. 5 as a function of the nanoparticle radius with solid thick (thin) inclined line. The most probable distance between a surface and a closest point of a nanoparticle is shown in Fig. 5 as a function of the nanoparticle radius with a dashed inclined line; it nearly coincides for the two cut-offs chosen.

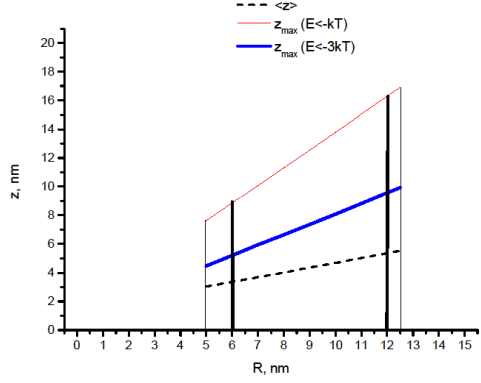

*Fig. 5. The range of characteristic parameters of the problem: the diamond nanoparticle radius $R[nm]$ on x-axis, and the spatial extension $z[nm]$ of quantum states on y-axis. The range of nanoparticle radii, corresponding to the efficiency of VUCN detection higher than 10% (1%) of the maximum efficiency is shown with vertical solid thick (thin) lines. The classical turning point for the nanoparticle with the energy $-3k_BT$ ($-k_BT$) in the bounding vdW/CP potential is shown as a function of the nanoparticle radius with solid thick (thin) inclined line. The most probable distance between surface and the closest point of the nanoparticle is indicated as a function of the nanoparticle radius with a dashed inclined line; it nearly coincides for the two cut-offs chosen.*

As clear from Fig. 5, the spatial extension of quantum states might be compatible, in the range of interest, to the nanoparticle size. This is the last condition in ref. (13) required to allow us to consider the interaction of UCN with "gas of freely levitating nanoparticles". Also note that nanoparticles could move quasi-freely along surface on a scale significantly larger than the characteristic size of the problem of ~10 nm. Scattering of the levitating nanoparticles on surface inhomogeneities, and on each other, results in their diffusion motion along surface. If the surface density of nanoparticles is large enough, they might form a kind of two-dimensional super-lattice due to their mutual repulsion and attraction, but this story is beyond the scope of the present article.

*4.2. Effect of the vdW/CP potential shape to VUCN spectra.*

At a later stage, developments of the study presented in this paper should allow one to use VUCN as probes for the potential well shape and to test details of the vdW/CP interaction. At certain conditions, a quasi-classical description used above could not be applied and we would observe purely quantum effects in the interaction of UCN with levitating nanoparticles. The conditions of strong contribution of precise shape of the vdW/CP potential to VUCN spectra include, for instance, very low temperatures (only lowest quantum states populated) and measurement of the momentum transfer; transitions between close quantum states at moderate temperature and measurement of the momentum transfer; three-dimensional surface potentials (due to surface corrugation) and measurement of the energy transfer; measurement of large energy transfer at moderate, or low, temperature on smaller nanoparticles, measurement of nanoparticle population in quantum states as a function of temperature etc. We will explore these and other cases, as well as methods of studying surface potentials using inelastic scattering of UCN on levitating nanoparticles, in further publications.

## 5. Comparison with experimental data

In this section we will compare all precision experimental data available on the VUCN production to calculations performed in framework of the simplest model of two-dimensional gas of freely levitating nanoparticles.

*5.1. Inelastic scattering of UCN on solid surfaces*

As the VUCN spectra have been measured for diamond nanoparticles rather precisely and our model calculates them with minor uncertainties, we will start comparison by these data sets. The sample studied here was a powder of diamond nanoparticles (53), (54). The radius of a typical single nanoparticle ranges from 1 nm to 10 nm; nanoparticles might agglomerate in clusters with larger effective radii and masses (55). As long as the cluster size is smaller than the UCN wavelength, precise shape of a nanoparticle or a cluster, as well as uniformity of their neutron-optical density, do not affect the VUCN spectra significantly. Also minor impurities do not change the resulting nuclear optical potential in a noticeable extend. Thus precise shape, density and chemical composition distributions could be ignored at this stage. In the following description, we will present real nanoparticles/clusters by idealized uniform spheres of equivalent effective size. Typical effective size distributions of diamond nanoparticles analogous to those used in ref. (30) are shown in Fig. 6; scattering of these distributions correspond to uncertainties in their precise knowledge related to the mentioned clustering effect.

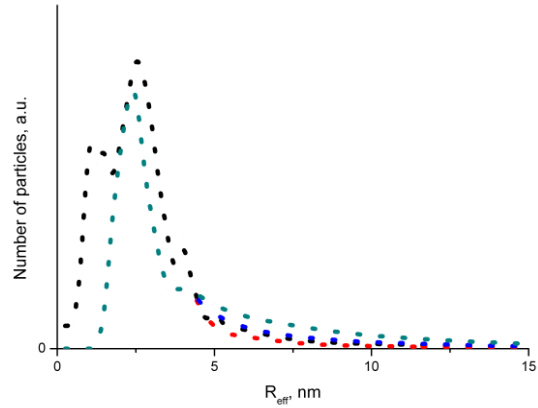

*Fig. 6. Distribution of the effective radii of detonation diamond nanoparticles measured using electron microscopy and X-ray scattering. The nanoparticles are assumed to be spherical and uniform in density; amorphous carbon-like shell and clustering of nanoparticles are taken into account. Different curves are related to various assumptions on nanoparticle clustering and to different measuring procedures.*

The VUCN spectra measured in ref. (30) using nano-diamond samples at ambient temperature and at the liquid nitrogen temperatures are given in Fig. 7a. A typical energy transfer is a few tens neV; saturation of the integral spectra at larger energies indicates significantly smaller probability of larger energy transfers; smaller temperature corresponds to smaller probability of VUCN detection rate; the measured raw data on the suppression factor in the VUCN detection rate is 1.9±0.2 that corresponds to the suppression factor in the total probability of VUCN production 2.2±0.2, taking into account increase in the VUCN detection efficiency at lower temperature.

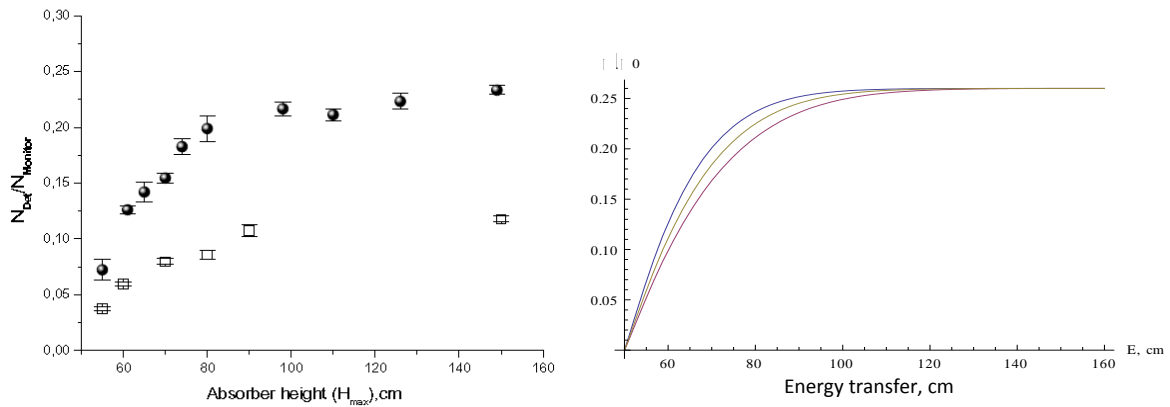

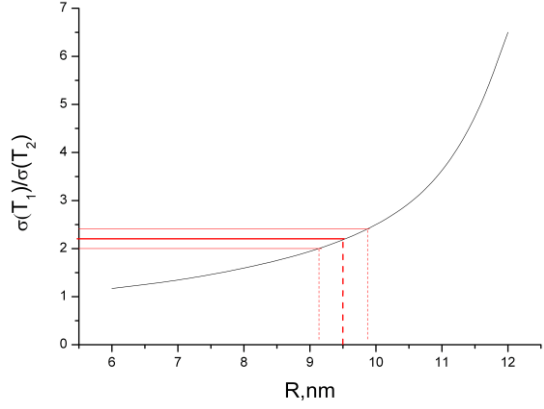

*Fig. 7.*

*a) Measurements with powder of diamond nanoparticles: the number of VUCN detected as a function of the maximum absorber height, normalized to the number of monitor counts (per main part of the measurement cycle). Black circles correspond to the measurement at ambient temperature. White squares indicate results of measurements at the temperature 108 K.*

*b) Theoretical simulations of the integral spectra of VUCN produced due to UCN inelastic scattering on mono-sized diamond nanoparticles with the radius (from top to bottom) equal 9.8 nm, 9.4 nm and 9.0 nm at ambient temperature. Y-axis indicates the probability of VUCN detection in the sensitivity window in the BGS spectrometer normalized to the total probability of inelastic scattering of UCN. All three curves are normalized to one value at the height of 160 cm, for illustrative purposes. Y-axis corresponds to the absolute probability of VUCN detection in the BGS spectrometer for a nanoparticle with the radius 9.4 nm.*

*c) The ratio of the VUCN count rate with the maximum absorber height of 150 cm calculated for the ambient temperature of 293 K to that calculated for the temperature of 108 K, as a function of the nanoparticle radius R, nm. The horizontal lines correspond to the measured ratio and the statistical confidence corridor for this value (Fig. 7a); systematic uncertainty of this value is twice larger. The vertical lines indicate the corresponding nanoparticle radius and the confidence range for this value.*

A simplified analysis could be performed assuming mono-sized nanoparticles with the radius $\bar{R}$. This assumption is justified at this stage by the resonance character of efficiency of VUCN detection in the BGS spectrometer (see Fig. 4) as a function of the nanoparticle radius. As clear from Fig. 7b, the calculated integral VUCN spectra are very sensitive to a precise value of the radius $\bar{R}$. The thermal velocity of larger nanoparticles is smaller, thus the energy transferred to UCNs is smaller, the integral spectrum is harder, and vice versa. This sharp dependence allows us to estimate, within the model of mono-sized nanoparticles, the effective nanoparticle radius corresponding to the data in Fig. 7a as well as uncertainties of this estimated value; the result is $\bar{R}_{diamond}^{(spectrum)} = 9.4 \pm 0.4 \, nm$.

Our model does not assume explicitly a change in the total probability of VUCN production on solid surfaces as a function of temperature, provided that the total number and size distribution of levitating nanoparticles is conserved that is definitely true around the ambient temperature, and below, as all processes of nanoparticle formation/ evolution are slower than a characteristic measurement time. However, even in this case, lower temperature of nanoparticles corresponds to smaller their velocities, thus to softer VUCN spectrum. As the range of initial UCN velocities is separated from the window of efficient VUCN detection by a dead-zone with zero, or small, probability of VUCN detection, a larger fraction of VUCN fit to the dead-zone at lower temperature, thus smaller fraction of VUCN is detected in the BGR spectrometer, as well as in any other spectrometer. This effect is shown in Fig. 7c. Using a measured value of the ratio of VUCN count rate at ambient temperature to that at the liquid-nitrogen temperature (corresponding to the data in Fig. 7a), we could estimate the effective nanoparticle radius: $\bar{R}_{diamond}^{(temperature)} = 9.5 \pm 0.6 \, nm$.

Although a fraction of relatively large nanoparticles in the realistic nano-particle size distribution in Fig. 6 is small, nevertheless they strongly contribute to the inelastic UCN scattering. Namely the contribution of particles/clusters with the radius of ~10nm is enhanced as clear from Fig. 5. For completeness, one should note that the sample of diamond nano-powder in ref. (30) was thick; therefore, strictly speaking, only a fraction of nanoparticles levitated above a copper surface, other nanoparticles interacted with each other, so one should consider also the interaction between diamond nanoparticles in the powder sample. However, as long as we stay within the model of free nanoparticles, this is not important; the effects following from the vdW/CP interaction with a substrate will be considered later. Convolution of the effective nanoparticle radii, shown in Fig. 6, with the efficiency of VUCN detection, illustrated in Fig. 4, results to the effective mean nanoparticle radius of $\bar{R}_{diamond}^{(average\ size)} = 8.5 \pm 0.5\ nm$; the uncertainty in $\bar{R}$ value corresponds to uncertainty in knowledge of actual nano-particle size distribution.

One should note that the agreement between these three independent estimations $\bar{R}_{diamond}^{(spectrum)} = 9.4 \pm 0.4\ nm$, $\bar{R}_{diamond}^{(temperature)} = 9.5 \pm 0.6\ nm$, and $\bar{R}_{diamond}^{(average\ size)} = 8.5 \pm 0.5\ nm$ is particularly convincing keeping in mind that the spectra shapes and the temperature dependence are not very sensitive to such model parameters such as precise size distribution of nanoparticles, surface roughness, impurities, shapes of nanoparticles, the effect of nanoparticle clustering etc. On the other hand, if you abandon the "free nanoparticle" model and "turn" artificially some critical parameter by only 10-20%, for instance the radius in eq. 3.21, without affecting any other parameter, the theoretical predictions would largely contradict the data. Strictly speaking, if you abandon the "free nanoparticle" model, there is no reason then why these values of $\bar{R}$ are equal, or even compatible in the order of magnitude.

The measured probability of VUCN production to the sensitivity window in the BGS spectrometer in ref. (30) was equal to ~$10^{-3}$ per one collision with the surface covered with nanoparticles. In order to estimate the absolute probability reliably, we would need to have carried out an experiment with a sample consisting of single diamond nanoparticles on a copper surface separated by distances much larger than a nanoparticle size. As this condition was not met in ref. (30), we could present only an approximate upper bound for this probability. We will assume that only nanoparticles above a copper surface and surface nanoparticles above powder (not in the powder bulk) could be weakly bound thus contribute to the measured effect; only nanoparticles with the optimum size of 8.0-9.4 nm could contribute; such nanoparticles should be separated by distances larger than their size in order to behave as freely levitating objects; VUCN produced in collisions with such nanoparticles should fit within the window of efficient detection in the BGS spectrometer. Multiplication of the mentioned probabilities gives an approximate upper bound of $(2-10) \cdot 10^{-3}$ in reasonable agreement to the measured value.

Another measured sample is nanoparticles appearing on a stainless steel surface due its thermal treatment (30). One should note that, in contrast to the previous measurement, in which we introduced ourselves diamond nanoparticles with a known size distribution, here we deal with naturally growing clusters due to atomic thermal diffusion on surface. As this measurement, for the sample pre-heating temperature of 350C, was performed with even better statistical precision than the measurement with diamond nanoparticles, the results allow us to calculate numerically a differential VUCN spectrum (by means of subtracting count rates in neighbor experimental points in the integral spectrum). The measured integral and differential spectra of VUCN are shown in Fig. 8a.

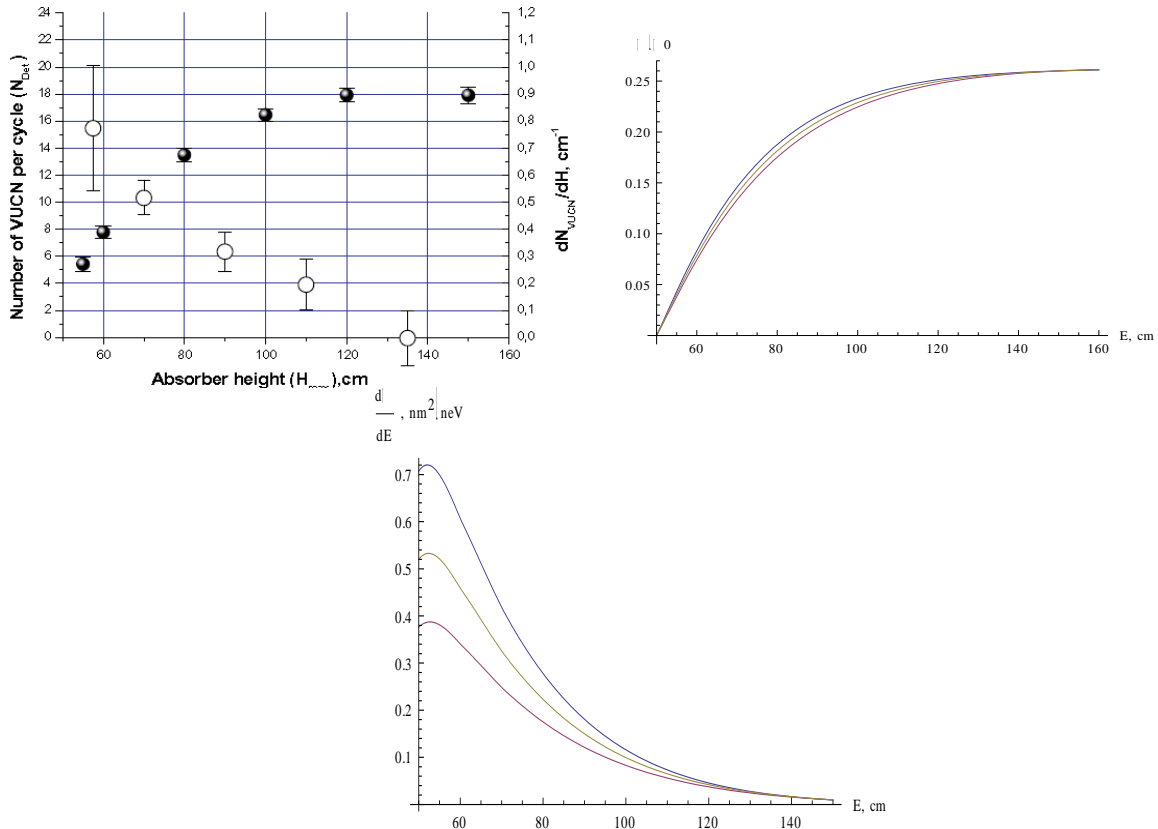

*Fig. 8.*

*a) Integral (left-hand ordinate axis, solid circles) and differential (right-hand ordinate axis, open circles) spectra of neutrons scattered inelastically on a stainless steel surface versus the height of the absorber upper position. The integral spectrum is the number of VUCNs detected per circle. The differential spectrum is obtained by numerical differentiation of the integral spectrum corrected to the efficiency of VUCN detection. The temperature of pre-heating the sample surface is 350 C.*

*b) Theoretical simulations of the integral spectra of VUCN produced due to UCN inelastic scattering on mono-sized nanoparticles with the volume density of stainless steel and with the radius (from top to bottom) 6.6 nm, 6.3 nm and 6.0 nm at ambient temperature. Y-axis indicates the probability of VUCN detection in the sensitivity window in the BGS spectrometer normalized to the total probability of inelastic scattering of UCN. All three curves are normalized to common value at the height of 160 cm, for illustrative purposes. Y-axis corresponds to the absolute probability of VUCN detection in the BGS spectrometer for a nanoparticle with the radius 7.0 nm.*

*c) Theoretical simulations of the corresponding differential spectra of VUCN produced due to UCN inelastic scattering at mono-sized nanoparticles with the volume density of stainless steel and with the radius (from top to bottom) 6.6 nm, 6.3 nm and 6.0 nm at ambient temperature.*

We do not know precise chemical (elementary) composition of nanoparticles formed on surface of stainless steel samples; therefore, as a starting point, we will assume that their density and nuclear optical potential are equal to those for stainless steel. If so, we could estimate the average nanoparticle radius from the measured integral and differential spectra in Fig. 8; it is equal $\bar{R}_{stainless\ steel}^{(spectrum)} = 6.3 \pm 0.3\ nm$.

The size distribution of nanoparticles observed in an AFM microscope on surface of stainless steel samples identical to that used in the neutron measurements is shown in Fig. 9. In analogy to the procedure used for the data analysis concerning diamond nanoparticles, convolution of the effective nanoparticle radii, shown in Fig. 9, with the efficiency of VUCN detection, illustrated in Fig. 4, results to the effective mean nanoparticle radius of $\bar{R}_{stainless\ steel}^{(average\ size)} = 6.6 \pm 0.3\ nm$; the uncertainty in $\bar{R}$ value corresponds to the uncertainty in measurements of actual nano-particle size distribution.

As for diamond nanoparticles, independent estimations of the value of effective average radius of nanoparticles $\bar{R}_{stainless\ steel}^{(spectrum)} = 6.3 \pm 0.3\ nm$ and $\bar{R}_{stainless\ steel}^{(average\ size)} = 6.6 \pm 0.3\ nm$ agree with each other, while the values themselves for diamond and stainless steel nanoparticles are different (eventually due to their different density). It is interesting to note that masses of diamond and stainless steel nanoparticles estimated above are equal to each other, thus providing an additional test of validity of our model.

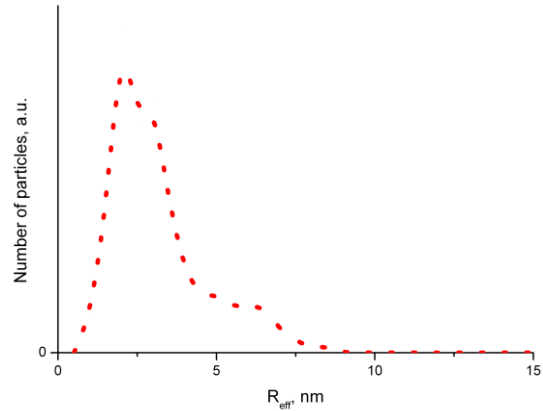

*Fig. 9. Distribution of effective radii of nanoparticles observed on surface of stainless steel samples pre-heated to the temperature of 350C; the measurement is performed at ambient temperature using an atomic-force-microscope.*

Note that the temperature dependence analogous to that for diamond nanoparticles and nanoparticles on stainless steel surface was observed in measurements with copper nanoparticles on a copper surface, as well as with beryllium nanoparticles on a beryllium surface (23), however these studies have not been completed thus could not be used for the present analysis.

A particularly sensitive another result is the observation of so-called "temperature resonance" shown in Fig. 10: sharp increase in the probability of inelastic UCN scattering on stainless steel surfaces as a function of the sample preceding heating temperature. We will apply our model to the experimental data and show that such "resonances" appear naturally in framework of the nanoparticle hypothesis. In this experiment we measured the probability of VUCN production on stainless steel surfaces as a function of the temperature of preceding heating of these stainless steel samples. The initial UCN spectrum and the spectrometer efficiency are analogous to those used in experiments with diamond nanoparticles. Additional measurements with an atomic-force microscope indicated intense formation of nanoparticles/nanostructures on a stainless steel surface, with the characteristic size equal approximately to the size corresponding to optimum sensitivity in the BGS spectrometer for VUCN;

such intense nanoparticle formation takes place precisely at the temperature of sharp increase in the VUCN production probability. A general feature of the nanoparticle formation is well pronounced: their size increases monotonously with increasing the heating temperature; nearly no particles at ambient temperature; intense growth of the number and size of nanoparticles at the temperature below ~350C; further increase in the size but decrease in the number of nanoparticles at the temperature higher than ~350C (some nanoparticles might coagulate).

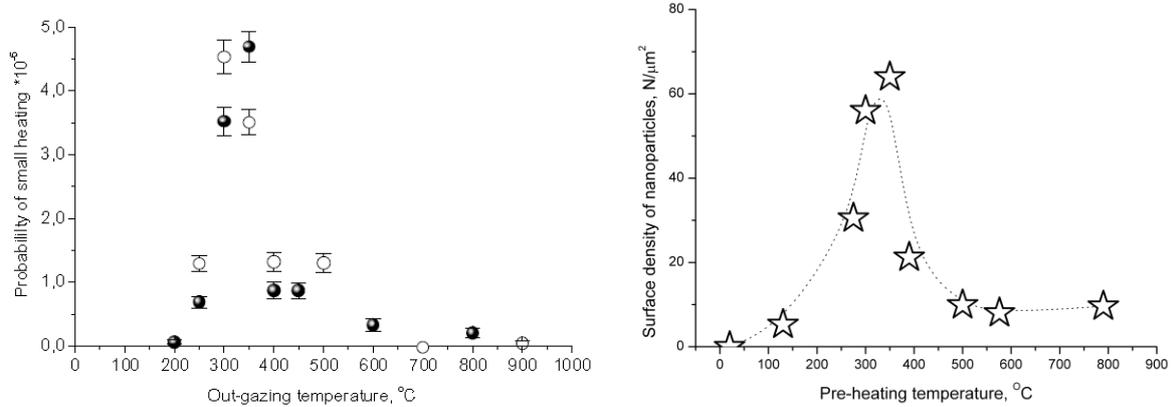

*Fig.10.*

*a) The probability of inelastic scattering of UCN on a surface of stainless steel samples is shown as a function of the temperature of sample outgasing (pre-heating). The measurement is carried out at ambient temperature. Black and white points show results of two independent measurements using analogous samples. The initial UCN spectrum is shaped to the range of 0-50 neV. The probability of inelastic scattering is restricted here to events of neutron scattering to the energy higher than 55neV in one wall collision. b) The surface density of nanostructures (at analogous stainless steel samples) with a diameter in the range of 12-14 nm is shown with stars; 1-σ error bars are approximately equal to their size; the dotted line is a b-spline of the data given for illustrative purposes.*

Note that the maxima in Figs. 10a,b coincide quite well, whereas analogous analysis carried out for not precisely optimum nanoparticle sizes in ref. (37) resulted to a peak in the figure analogous to Fig. 10b shifted by about 50C towards lower temperature. This observation provides another indication that our model describes the existing experimental data quite in detail.

Finally, nor nanoparticles neither inelastic scattering of UCN are observed on crystal sapphire surface (27), (30) thus providing one more test of validity of our model.

To summarize: our model describes quantitatively spectra and temperature dependences for all the experimental data available for VUCN production on solid surfaces with the accuracy of the order or 10% or so in terms of the particle sizes. There are no alternative interpretations of these experimental data available; in particular, as shown in ref. (31), thermal fluctuations of solid surfaces provide negligible effects.

*5.2. Inelastic scattering of UCN on liquid surfaces.*

Another data set on inelastic scattering of UCN is related to experiments with liquid surfaces, in particular with Fomblin oils. We assume that inelastic scattering of UCN on liquid surfaces is, at least partly, due to collisions of UCN with Fomblin oil nano-droplets, with relatively broad size distribution of the nano-droplets, trapped via the mechanism described above; it complements the known effects due to surface thermal fluctuations (31) and surface capillary waves (56), and plays the major role for the experiment considered here. Distinctive features of our model consist of relatively large mean energy transfers, as well as of a different temperature dependence of the VUCN production rates. As a result of these and other effects, it could be easily verified in dedicated experiments. Inelastic scattering of UCN on liquid oil Fomblin surfaces has been studied poorly in spite of the fact that it was discovered earlier (15), (18), (16) than inelastic scattering of UCN on solid surfaces, also in spite of the fact that it was measured in many experiments performed by various groups. The only parameter actually measured in the preceding experiments is the total probability of inelastic scattering of UCN from some broad initial energy range to another broad final energy range. These data are irrelevant for our purposes of comparing the theory to experimental data. Also reliable data on the temperature dependence of VUCN differential spectra are missing. On the other hand, scattering of this kind is relatively easy to observe due to very large scattering probability, soft VUCN spectra thus high detection efficiency. That is why we are going to perform a dedicated experimental study. Now, we will predict the integral and differential spectra, as well as the detection rate of VUCN produced on Fomblin oil nano-droplets as a function of temperature; the spectrometer parameters are assumed to be equivalent to those used in measurements with solid nanoparticles. Data for other spectral parameters of other eventual experiments could be easily estimated using the method used above. The optimum sensitivity in the BGS spectrometer for Fomblin oil nano-droplets is shown in Fig. 11a. As the density of Fomblin oil is not much different from that of diamond, the sensitivity curve is similar to that shown in Fig. 4 for diamond. As the size distribution of Fomblin oil nano-droplets is broad, the effective radius is close to $\bar{R}_{Fomblin\ oil}^{(size\ distribution)} = 10\ nm$, which corresponds to the maximum value in Fig. 11a. The total probability of VUCN detection in the BGS spectrometer is shown in Fig. 11b as a function of temperature; the total number of levitating nanoparticles is assumed to be independent on temperature in this figure. The later condition is not met for liquid droplets, it is rather proportional to the vapor pressure of Fomblin oil above surface; anyway this dependence has to be studied in detail experimentally and theoretically. In contrast to solid nanoparticles with their very long characteristic times of formation and evolution, the number and size of levitating liquid nano-droplets could rapidly change with temperature, or other parameter, as they result from equilibrium between two processes: permanent formation of nano-droplets from vapors, also from explosion of bubbles on surface, and their evaporation. The integral and differential VUCN spectra are simulated in Figs. 11c,d. As the calculated spectra are rather soft, it would be probably more convenient to shape the initial UCN spectrum in actual experiments differently, for instance, from zero to 25 neV, or from zero to 30 neV.

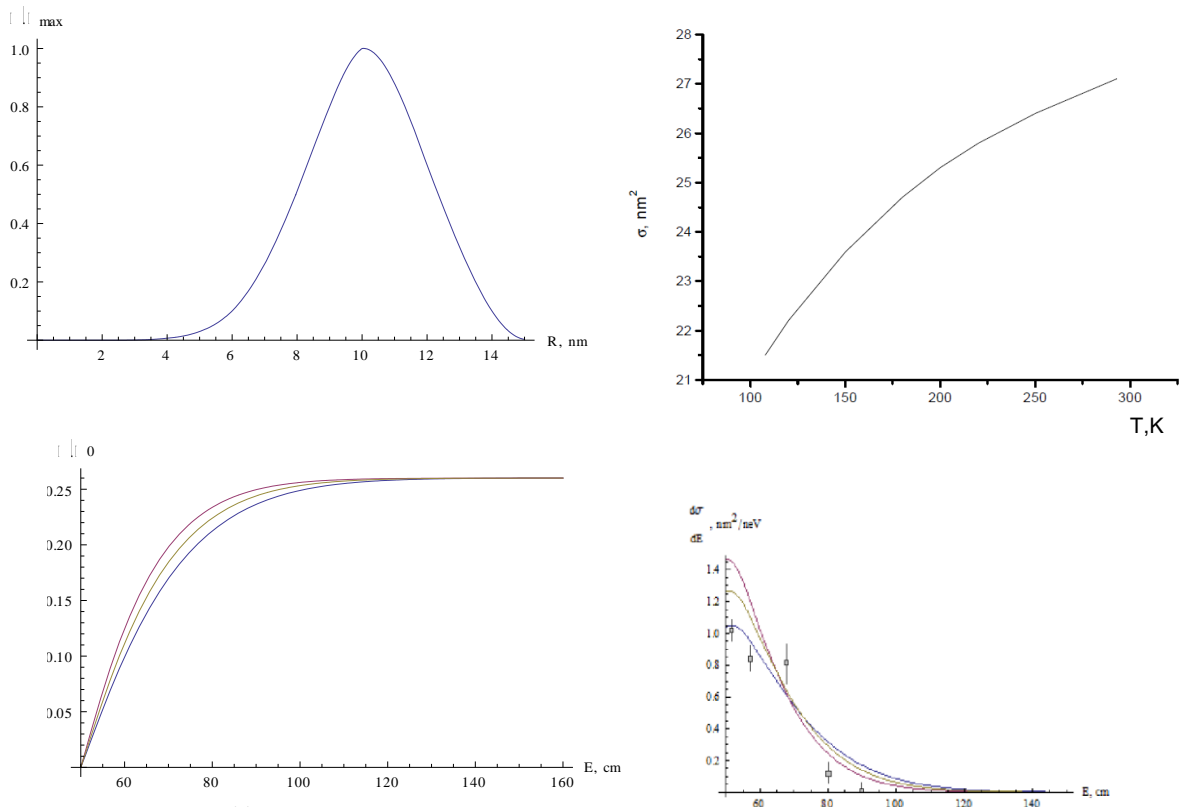

Fig. 11.
a) *The total probability of VUCN detection in the BGS spectrometer (the spectrometer efficiency $Eff(E)$ is taken into account) as a function of the Fomblin oil nano-droplet radius $R[nm]$.*
b) *The cross section of VUCN inelastic scattering to the sensitivity window in the BGS spectrometer (the spectrometer efficiency $Eff(E)$ is taken into account; the number of nano-droplets is assumed for simplicity to be constant in this calculation) as a function of the temperature $T$; the nano-droplet radius is 10 nm. A difference between this prediction and experimental data would indicate on changes in the number of nano-droplets of optimum size as a function of temperature.*
c) *Theoretical simulations of the integral spectra of VUCN produced due to UCN inelastic scattering at mono-sized Fomblin oil nano-droplets with the radius (from top to bottom) 10.4 nm, 10.0 nm and 9.6 nm at ambient temperature. Y-axis indicates the probability of VUCN detection in the sensitivity window in the BGS spectrometer normalized to the total probability of inelastic scattering of UCN. All three curves are normalized to a common value at the height of 160 cm for illustrative purposes. Y-axis corresponds to the absolute probability of VUCN detection in the BGS spectrometer for a nanoparticle with the radius 10.0 nm.*
d) *Theoretical simulations of the corresponding differential spectra of VUCN produced due to inelastic scattering of UCN at mono-sized Fomblin oil nano-droplets with the radius (from top to bottom) 10.4 nm, 10.0 nm and 9.6 nm at ambient temperature. Preliminary experimental data, measured by V.I. Morozov's group, are indicated with open boxes.*

## 6. Manifestations and applications of inelastic scattering of UCN on levitating nanoparticles

Careful analysis of various manifestations and applications of inelastic scattering of UCN on levitating nanoparticles will be performed later; here, we only list them briefly.

We have shown in the present work that the observed small-energy-transfer inelastic scattering of UCN on solid surfaces (16), (17), (19), (23), (26), (27) (30) is explained by their scattering on levitating nanoparticles above surface. Within the presented model of two-dimensional gas of freely levitating nanoparticles, we have described all the existing reliable experimental data. Note that our model provides the only explanation to the observed phenomenon as the thermal fluctuations of surface (31) are negligible for the measured VUCN production on solid surfaces.

Inelastic scattering of UCN on beryllium nanoparticles levitating above beryllium surface might contribute to so-called anomalous losses of UCN (38) only if the probability of such inelastic scattering is compatible to the "anomalous" probability of $1.3 \cdot 10^{-5}$ per collision. As measured in refs. (19), (23), the probability of UCN inelastic scattering on a beryllium surface to the window of efficient detection in the BGS spectrometer was lower but compatible to that on stainless steel (~$10^{-7}$ at ambient temperature, no surface treatment was applied), and the temperature damping of VUCN detection probability is approximately equal for stainless steel, diamond, copper and beryllium surfaces (thus masses of nanoparticles of all these materials and VUCN spectra are compatible). However, stainless steel surface out-gazing at the temperature 350 C increased the probability by about two orders of magnitude. In order to reveal relevance of inelastic scattering of UCN on nanoparticles to the anomalous UCN losses, we will study formation of nanoparticles with the optimum characteristic radius as a function of surface heating, and, later if needed we will measure directly VUCN production at equivalent conditions. It is curious to mention that the value of the "temperature resonance" considered in the present article and observed in refs. (30), (37), is equal to the typical temperatures 300-400C of heating/outgasing of storage volumes in the neutron lifetime experiments needed to get rid of hydrogen contaminations on surface of UCN storage volumes. Thus when eliminating hydrogen with its large up-scattering cross-section from surfaces, one might introduce instead nanoparticles with their large cross sections of inelastic scattering of UCN. Note that negative results in ref. (29) are irrelevant to the present problem because of too low energy resolution of a spectrometer user there, too low VUCN detection efficiency and no thermal treatment of the samples used.

We argue that analogous inelastic scattering of UCN on Fomblin oil liquid surfaces (15), (18), (16), (17), (19), (24), (25), (28), (29) is explained in significant extend by their scattering on levitating Fomblin oil nano-droplets, complementary to the known inelastic scattering of UCN on surface capillary waves (56), and on surface thermal fluctuations (31). The model of inelastic scattering of UCN on levitating nanoparticles suggests harder differential spectra and significantly different temperature dependences of VUCN detection. We plan performing promptly dedicated experiments to verify our predictions.

The present study is motivated in part by unsatisfactory situation in the neutron lifetime measurements: the large and unexplained so far discrepancy between the neutron lifetime experiments with smallest reported uncertainties (40). This contradiction has been discussed concerning far-reaching consequences of eventual shift of the mean world value of the neutron lifetime for the fundamental particle physics and cosmology, but a little independent work has been done to verify validity of the measured results; a notable exception is ref. (57). All these experiments use storage of UCN in traps with Fomblin oil wall coatings and they rely on assumption of conservation of energy of UCN during their storage. Unfortunately this condition is not met thus leading to major false effects. As an example, we consider an effect of inelastic scattering of UCN on levitating Fomblin oil nano-droplets in the experiment (41). Measured probability of VUCN detection decreases, as a function of temperature, in analogy to the effect shown in Fig. 7a,c (measurement and model calculation). However, in contrast to very small dead-zone between the maximum initial UCN energy and minimum VUCN energy in the sensitive window in the BGS spectrometer, a dead-zone in the gravitational spectrometer "Kovsh" in ref. (58), (59), (41) as well as a dead-zone in the spectrometer used in ref. (29) is much broader, thus the temperature damping is stronger. The measured temperature damping (29) is shown in Fig. 12a. Our estimation of the cross-section of inelastic scattering of UCN on Fomblin oil nano-droplets is shown in Fig. 12b; we assume for simplicity here that the total number of nano-droplets does not depend on temperature. We cannot introduce at this stage an explicit dependence of the number of nano-droplets, as a function of temperature, as this requires direct experimental measurements or a theoretical description of nano-droplet formation.

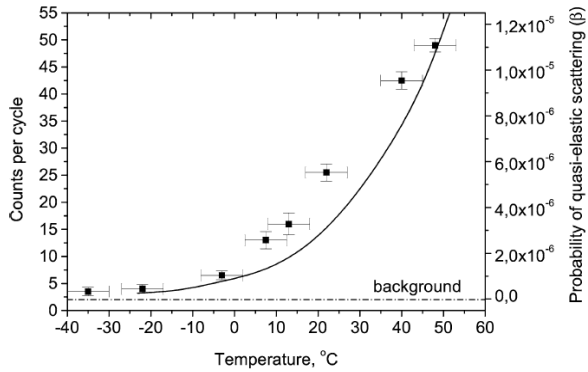 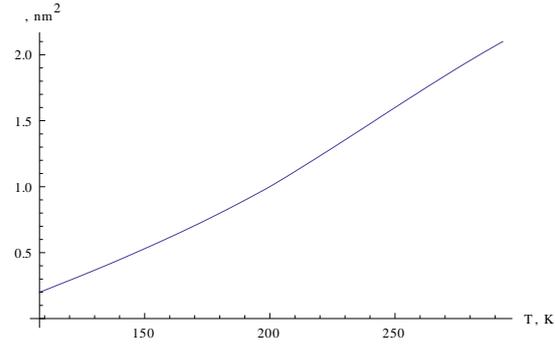

*Fig. 12. a) This figure is copied from ref.* (29). *The caption in ref.* (29) *says: "Temperature dependence of UCN low energy heating by Fomblin. On the left vertical scale, the integral counts in detector from 380<t<780 s before background subtraction are plotted. On the right vertical scale, the probability of the quasi-elastic heating is indicated. The solid curve is taken from ref.* (56); *it was calculated for capillary waves on the surface of Fomblin with initial UCN energies between 0 and 52 neV and final energies between 52 and 106 neV. The data and the calculation are qualitatively in good agreement." We will comment bellow on validity of this caption.*

*b) The cross-section of the inelastic scattering of UCN on a Fomblin oil nano-droplet with an "optimum" radius of 10 nm. Here, the total number of nano-droplets is assumed to be independent on temperature. The initial UCN spectrum and the range of final UCN energies that could be detected are taken from ref.* (29).

If the equilibrium surface density of nano-droplets is approximately proportional to the Fomblin oil vapor pressure above surface, our model dependence would describe the data (note this factor is not known at this stage). On the other hand, one might notice that the model of surface capillary waves (56) is not properly applied to the data in Fig. 12a, as a large dead-zone between the initial UCN energies and the final UCN energies is not taken into account. If you do so (see Fig. 13 in ref. (32)), the model dependence would even change the sign, as a function of temperature, and the total probability would be too low to explain the data.

If the nano-droplet hypothesis will appear to be correct, all neutron lifetime experiments should be reanalyzed and corrected if needed. Inelastic scattering is "invisible" because of poor spectral sensitivity of the spectrometers used; "visible" VUCN is a fraction of the "iceberg". It is easy to estimate, keeping in mind typical storage times, the frequency of UCN-wall collisions, and the total probability of quasi-elastic scattering of UCN on a Fomblin oil surface that the probability of a change in energy for every UCN is significant. UCN do change their energy thus their loss rate is different from the calculated values and false effects appear. Most unpleasant one might follow from the fact the rate of forming Fomblin oil nano-droplets depends on Fomblin oil vapor pressure above surface and amount and composition of gases diluted in Fomblin oil; it changes in time and depends somehow on "weather" above surface…. Thus the geometrical method of size extrapolation, as well as the method of energy extrapolation, could not be applied to precision neutron lifetime experiments without reservations. In particular, as a result the loss probability would be different in large and small traps used for the size extrapolation. The task of estimating corrections to the neutron lifetime values in various experiments goes beyond the scope of the present article.

In order to complete the list of manifestations and applications of inelastic scattering of UCN on levitating nanoparticles we note that a particularly attractive application of the phenomenon presented in this article might consist of studying the vdW/CP interaction between levitating nanoparticles and surface. Besides, complementary to using of trapped atomic clusters for decorating surface defects, boundaries, step edges, grain boundaries, elastic strain fields (see, for instance, (60), (61), (62)), UCN are sensitive to nanoparticles/ atomic clusters in the state of motion above defect-free zones, thus giving us access to nanoparticle mobility. Due to levitation of nanoparticles above surface, their mobility is very high (61); on the other hand chemical interactions (63) largely reduce it thus giving us access to chemical properties of nanoparticles and surfaces. In analogous way, one could measure electrostatic effects.

## 7. Conclusions

We argue that nanoparticles with sizes from a few nanometers to a few tens of nanometers might levitate above surface due to their physical adsorption in the potential well formed by the attractive vdW/CP forces on one side and quantum reflection from a surface on another side. The depth of this potential well is much larger than the thermal energy. At certain conditions, often met in realistic experiments, such objects behave as two-dimensional gas of quasi-freely levitating nanoparticles.

We describe quantitatively such levitating nanoparticles as well as the interaction of UCN with them. We argue that small-energy-transfer inelastic scattering of UCN on solid surfaces, so-called small heating of UCN, is due to their scattering on nanoparticles adsorbed to surface. Our model suggests very robust simulation of the data, which could be, or have been, measured. In particular, a kind of "signature" for such inelastic scattering of UCN on stainless steel surfaces, and probably on many other surfaces as well, is a sharp "resonance" increase in the inelastic scattering probability as a function of the temperature of surface preceding outgasing. We show that such "resonance" appears naturally within our model. Moreover, it describes the temperature dependence of VUCN detection, the integral and differential spectra, as well as the measured size distribution of nanoparticles in a self-consistent manner.

We assume that VUCN on liquid surfaces could be produced on levitating nano-droplets. We predict VUCN spectra at various temperatures that might be observed in bottles with Fomblin oil liquid surfaces; we are going to verify our predictions in dedicated experiments. If our expectations were confirmed, this effect causes major false effects in precision neutron lifetime experiments.

A particularly attractive future application of the inelastic scattering of UCN on levitating nanoparticles might consist of studying the vdW/CP interaction between the levitating nanoparticles and surfaces; if so, the range of accessible materials and temperatures is extremely broad and could include very exotic cases.